\documentclass[acmsmall]{acmart}
\usepackage{todonotes}
\usepackage{subcaption}
\newcommand{\hl}[1]{#1}
\AtBeginDocument{%
  \providecommand\BibTeX{{%
    \normalfont B\kern-0.5em{\scshape i\kern-0.25em b}\kern-0.8em\TeX}}}

\setcopyright{acmlicensed}
\acmJournal{PACMHCI}
\acmYear{2023} \acmVolume{7} \acmNumber{ETRA} \acmArticle{158} \acmMonth{5} \acmPrice{15.00}\acmDOI{10.1145/3591127}

\begin{document}
\title[DynamicRead on Mobile]{DynamicRead: Exploring Robust Gaze Interaction Methods for Reading on Handheld Mobile Devices under Dynamic Conditions}





\author{Yaxiong Lei}
\email{yl212@st-andrews.ac.uk}
\orcid{0000-0002-0697-7942}
\affiliation{%
  \institution{University of St Andrews}
  \city{St Andrews}
  \country{UK}
  \postcode{KY16 9SX}
}

\author{Yuheng Wang}
\email{yw99@st-andrews.ac.uk}
\orcid{0000-0003-3335-8706}
\affiliation{%
  \institution{University of St Andrews}
  \city{St Andrews}
  \country{UK}
  \postcode{KY16 9SX}
}

\author{Tyler Caslin}
\email{tc80@st-andrews.ac.uk}
\orcid{0009-0008-0158-2563}
\affiliation{%
  \institution{University of St Andrews}
  \city{St Andrews}
  \country{UK}
  \postcode{KY16 9SX}
}

\author{Alexander Wisowaty}
\email{arkw1@st-andrews.ac.uk}
\orcid{0009-0004-6660-6512}
\affiliation{%
  \institution{University of St Andrews}
  \city{St Andrews}
  \country{UK}
  \postcode{KY16 9SX}
}

\author{Xu Zhu}
\email{xz32@st-andrews.ac.uk}
\orcid{0000-0002-2801-3271}
\affiliation{%
  \institution{University of St Andrews}
  \city{St Andrews}
  \country{UK}
  \postcode{KY16 9SX}
}

\author{Mohamed Khamis}
\email{Mohamed.Khamis@glasgow.ac.uk}
\orcid{0000-0001-7051-5200}
\affiliation{%
 \institution{University of Glasgow}
 \city{Glasgow}
 \country{UK}}
 
\author{Juan Ye}
\email{Juan.Ye@st-andrews.ac.uk}
\orcid{0000-0002-2838-6836}
\affiliation{%
 \institution{University of St Andrews}
  \city{St Andrews}
  \country{UK}
  \postcode{KY16 9SX}
}

\renewcommand{\shortauthors}{Lei et al.}

\begin{abstract}
 
Enabling gaze interaction in real-time on handheld mobile devices has attracted significant attention in recent years. An increasing number of research projects have focused on sophisticated appearance-based deep learning models to enhance the precision of gaze estimation on smartphones. This inspires important research questions, including how the gaze can be used in a real-time application, and what type of gaze interaction methods are preferable under dynamic conditions in terms of both user acceptance and delivering reliable performance. To address these questions, we design four types of gaze scrolling techniques: three explicit technique based on Gaze Gesture, Dwell time, and Pursuit; and one implicit technique based on reading speed to support touch-free, page-scrolling on a reading application. We conduct a 20-participant user study under both sitting and walking settings and our results reveal that Gaze Gesture and Dwell time-based interfaces are more robust while walking and Gaze Gesture has achieved consistently good scores on usability while not causing high cognitive workload.
\end{abstract}

\ccsdesc[500]{Human-centered computing~Interactive systems and tools}
\ccsdesc[500]{Human-centered computing~Interaction techniques}
\ccsdesc[500]{Human-centered computing~Empirical studies in HCI}

\keywords{Eye Tracking, Mobile Devices, Smartphones, Gaze-based Interaction, Dwell, Pursuit, Gaze Gesture, Scrolling Techniques, Reading}

\maketitle
\section{Introduction}

In recent years, we have witnessed a significant improvement in the quality of cameras on our smartphones. This has attracted increasing attention to appearance-based gaze research on handheld mobile devices. This paper aims to understand and assess to what degree state-of-the-art gaze estimation techniques can support real-time applications. To this end, we design four gaze user interfaces for controlling page scrolling actions on a reading application.

Reading on a handheld mobile device is an everyday scenario~\cite{appannie2022mobile}. The average time people spend on their mobile phones is around 4.8 hours, and 7 out of every 10 minutes are spent on news, social media, and other similar apps for content reading.
Scrolling with touch is often needed for continuing reading longer articles or consuming news feed. As the screen size of smartphones grows, it becomes cumbersome to perform touch-based scrolling, especially in a single-handed use scenario where one hand is occupied with other tasks; for example, holding coffee. Thus, we rely on the other hand holding the phone and scrolling at the same time. However, in this case, the hand might not be able to reach the right scrolling region easily or may accidentally activate some other action. 

To support touch-free scrolling, we propose to use various gaze interaction methods to enable and control scrolling. Gaze-based scrolling on reading applications has been studied for as long as eye movement data have been deemed valuable in reading and other information-processing tasks~\cite{sharmin2013reading}. Various projects have experimented with off-the-shelf eye-tracking devices such as Tobii T60 and designed automatic scrolling techniques for reading on desktop screens~\cite{kumar2007gazeenhanced, sharmin2013reading, wilson2018autopager}. 

Our gaze user interface design is built on these existing projects and the main difference from them is that we use cameras in smartphones for gaze estimation and evaluate the usability of gaze interaction methods in real-time in mobile settings. The key challenge is that the spatial relation (i.e., distance and angle) between a user and the screen of their smartphones is constantly changing with different holding postures and motion states. This dynamic factor has presented as a barrier to adopting appearance-based gaze interfaces on smartphones~\cite{khamis2018understanding, khamis2018thepast}. 

To assess the usability of gaze interaction methods on real-time mobile applications, we design and develop a variety of gaze scrolling techniques, including implicit gaze where automatic scrolling is dynamically adapted to real-time reading speed, and explicit gaze where scrolling is explicitly activated by intentional gaze actions, i.e. Dwell, Pursuit and Gesture. We aim to explore the research questions, \textbf{RQ1:} How different types of gaze interaction methods work in real-time and in different scenarios, with focus on how human motion impacts gaze estimation, and \textbf{RQ2:} What type of gaze interaction methods is more robust while a user is walking. To the best of our knowledge, we are the first to evaluate the usability of various types of gaze interaction methods as scrolling technique in real-time mobile applications under both sitting and walking conditions. The user study results have shed insight on gaze interface design on handheld mobile devices. We also identify the pitfalls where gaze does not work well, which opens future research directions. The main contributions of this paper are summarised as follows: 
\begin{itemize}
    \item We design and develop four gaze scrolling techniques to enable touch-free scrolling and develop a reading application to support the full pipeline from collecting images from camera, to estimating gaze, and to activating gaze interfaces.
    \item We conduct a 20-participant within-subjects user study and systematically evaluate four types of gaze interaction methods (Dwell, Pursuit, Gesture and reading speed estimation) and compare with the touch-based method for reading under sitting and walking settings. The \href{https://doi.org/10.5281/zenodo.7806945}{eye movement dataset} is available for further investigation. 
    \item We analyse the results and uncover the new possibility of using gaze in mobile applications. 
\end{itemize}

\section{Related Work}\label{sec:relatedwork}
This section reviews the recent research appearance-based gaze estimation and gaze-supported scrolling techniques, both of which form the foundation of our work. We also look into the recent work on assessing gaze estimation under dynamic conditions. 

\subsection{Gaze Estimation on Smartphone}
Gaze interaction is to make use of gaze to facilitate interactions with computing devices. Gaze refers to a point on a screen or a direction in space and can be inferred from pupil positions, facial structure, and head movements. Benefiting from high-resolution cameras and high-performance processors, the appearance-based approach makes it easy to provide eye-tracking functionality in smartphones. In recent years,
many deep learning architectures are proposed for appearance-based gaze estimation and the performance on both 2D and 3D tasks has achieved significant progress~\cite{ghosh2021automatic, cheng2021appearance}. Convolutional neural networks (CNN) and convolutional layers are the most commonly used techniques in appearance-based methods to obtain facial and eye features~\cite{krafka2016eyetracking, palmero2018recurrent, zhang2018training, zhang2019evaluation, chen2018appearance, guo2019generalized, cheng2020coarse, bao2021adaptive, cheng2021gaze}.
Krafka et al. collected a large in-the-wild gaze dataset on mobile devices, \textit{GazeCapture}, and proposed a deep-learning based model, \textit{iTracker}. iTracker is a multi-branch CNN model, which takes images of a face, eyes, along with a face grid as input. It reaches a prediction error of 1.86cm and 2.81cm on iPhones and iPad respectively~\cite{cheng2021appearance}. For each of the three images, there are CNN branches to learn the features, and each branch consists of several convolutional layers. Because of its promising results, iTracker and its variants have been adopted in many gaze-interaction applications~\cite{krafka2016eyetracking}. GoogleGaze~\cite{valliappan2020accelerating} is another popular CNN model that takes images of the left eye and right eye along with the landmark positions of eye corners.  Similar to iTracker, GoogleGaze also employs one convolutional branch for each image for visual feature learning and then concatenates features on landmarks for final gaze point estimation. On Pixel 2 XL smartphones, GoogleGaze can reach the prediction error of 1.92cm and 0.54cm after 100-frame calibration.

\subsection{Gaze-based Scrolling Techniques}
Gaze-based scrolling techniques have been extensively investigated with off-the-shelf eye tracking devices. Kumar et al.~\cite{kumar2007gazeenhanced} have designed four types of scrolling techniques: \textit{smooth scrolling} that uses explicit gaze interactions to activate automatic scrolling and the activation is triggered by the dwell time of gaze in pre-defined regions; \textit{eye in the middle} that dynamically adjusts scrolling rate based on the estimated reading speed so that the users' gaze is kept at the middle of the screen; \textit{gaze-enhanced page up and page down} that highlight the region where the user is reading when the user presses the page down key; \textit{discrete scrolling} that uses dwell time to activate page-down command. Their results show that the participants prefer page-by-page scrolling rather than reading moving text. This work has inspired our explicit gaze-based scrolling techniques.


Sharmin et al.~\cite{sharmin2013reading,raiha2014gazescrolling} have designed an automatic scrolling technique in order to find preferred reading regions. Their experiment is conducted with Tobii T60 remote eye-tracking devices on a 17-inch monitor. Through a user study they investigate the preferred reading region on the screen and the impact of the font size on eye movement. Turner et al.~\cite{turner2015understanding} have proposed a gaze-enhanced scrolling technique that is based on the observation on manual scrolling. The observation can inform the design on automatically scrolling to make sure that the user is reading the content in their preferred reading region. Wilson and Williams~\cite{wilson2018autopager} have introduced ``autopager'' to render the unread text in the periphery so that viewers can continuously read without manual scrolling. A page is divided into the upper and lower section and as the user is reading one section, the other section will be presented at the periphery, allowing for continuous scrolling. The render automatically initiates based on the readers' current reading speed estimated from their gaze scan path. 


Our gaze scrolling techniques are inspired by the findings in the above work in terms of what gaze interfaces can be useful for scrolling. Our work is the first one to study real-time gaze-based scrolling using cameras of smartphones and appearance-based gaze estimation. For example, Autopager~\cite{wilson2018autopager} estimates gaze positions via a commercial eye tracker and renders text on a 24" screen. We assess the usability of these gaze interfaces under both sitting and walking conditions, which can uncover different reading patterns.


\subsection{Gaze Interaction Under Dynamic Conditions}
Mobility has a significant impact on the accuracy of gaze estimation. Most of the existing research focuses on the stationary conditions, where a user interacts with gaze while sitting or standing, and the example projects are FGFlick~\cite{yamato2021fgflick}, GAVIN~\cite{khan2021gavin}, and EyeSayCorrect~\cite{zhao2022eyesaycorrect}. WorldGaze~\cite{mayer2020worldgaze} is the only project that has applied both rear and front cameras for gaze-voice interaction and evaluated it under both stationary and dynamic conditions. However, their focus is on coarse-grained gaze direction estimation using head vector, rather than fine-grained on-screen point estimation.

\section{System Design}\label{sec:design}
This section illustrates the design and implementation details of our gaze interfaces, gaze estimation model and the reading application. 

\subsection{Gaze Reading Interface}
Our main objective is to evaluate the usability of scrolling techniques in real-time on a reading mobile application. We design four types of gaze scrolling techniques to demonstrate typical ways of utilising gaze, including gesture, dwell time, and pursuit. We focus on the scrolling functionality in a reading application, which represents the main behaviour on smartphones.  

Gaze-based scrolling has been extensively studied with off-the-shelf eye-tracking devices in a desktop setting. Our design is built on top of these studies~\cite{kumar2007gazeenhanced}. Firstly, we opt for page-by-page scrolling rather than continuous scrolling as the study has shown that readers find it disconcerting to read moving text. Secondly, instead of completely turning to a new page, we perform 95\% turning; that is, only 95\% of the page will be updated and the last line of the previous page will move to the top with an indication arrow. This prevents sudden scrolling, where readers might need to scroll back to check their last reading. In the following, we will introduce four types of gaze interfaces for page scrolling. 

\begin{figure}[!htbp]
\centering
\includegraphics[width=0.58\textwidth]{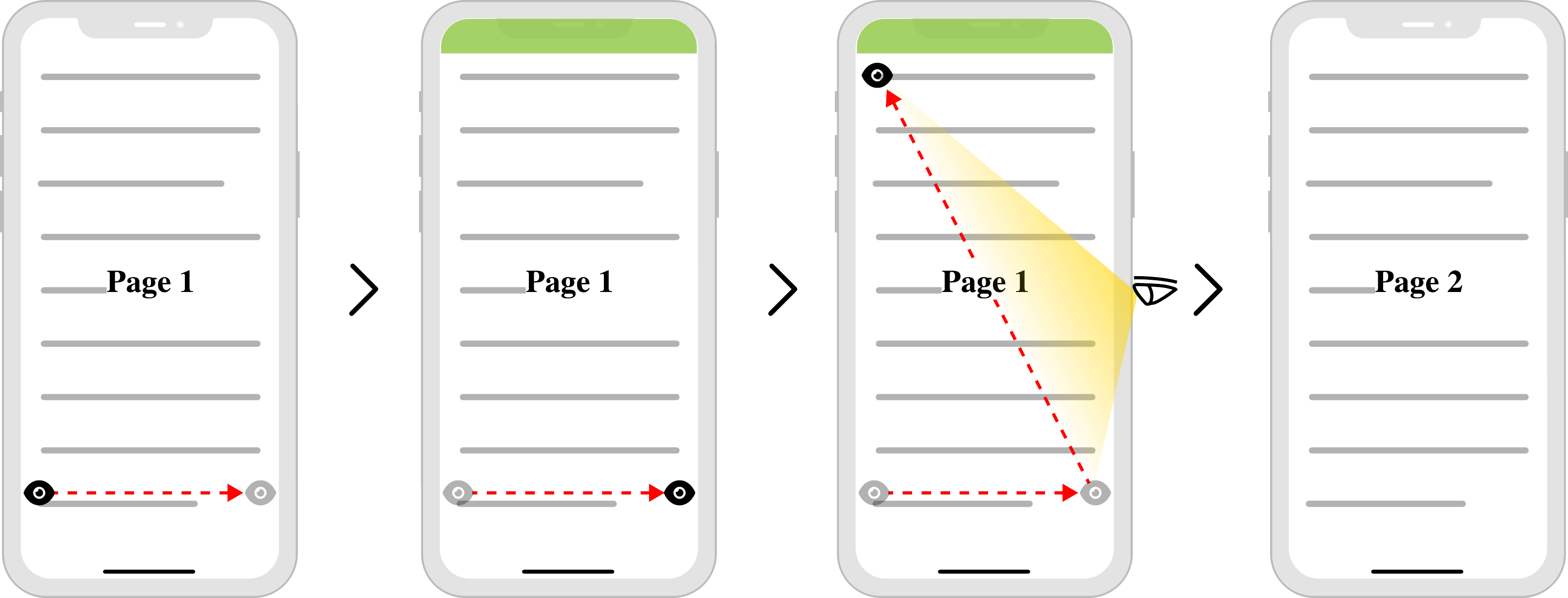}
\caption{Eye-Swipe: Gaze Gesture-based explicit gaze scrolling technique that a vertical gaze gesture moving from the bottom to the top of a page, simulating natural reading behaviours}
\label{fig:GazeC}
\end{figure}

\textit{Eye-Swipe} is an explicit gaze gesture-based scrolling technique where the gesture is a vertical transition from the bottom to the top of a screen, shown in Fig~\ref{fig:GazeC}. This gesture mimics the natural reading behaviour; that is, when we complete reading one page of content, our eye naturally comes to the beginning of the second page. The different stages of the gaze gesture can be abstracted as a finite-state machine. While the user is reading the text, they are in the reading state. Then, if the user dwells at the bottom of the screen – in other words if the y-position of the gaze estimation remains greater than 690 pixels (the last two lines) for at least 500 milliseconds, the user is then in the primed state. This step is makes \textit{Eye-Swipe} more robust to accidental eye movements. At this stage, the user can either look back to the text and keep reading, or move their gaze from the bottom to the top, indicating the final stage is reached. Then the colour of the top area turns green and the page is scrolled.

\begin{figure*}[!htbp]
\centering
\begin{subfigure}{0.3\textwidth}
\centering
\includegraphics[width=3.5cm, height=3cm]{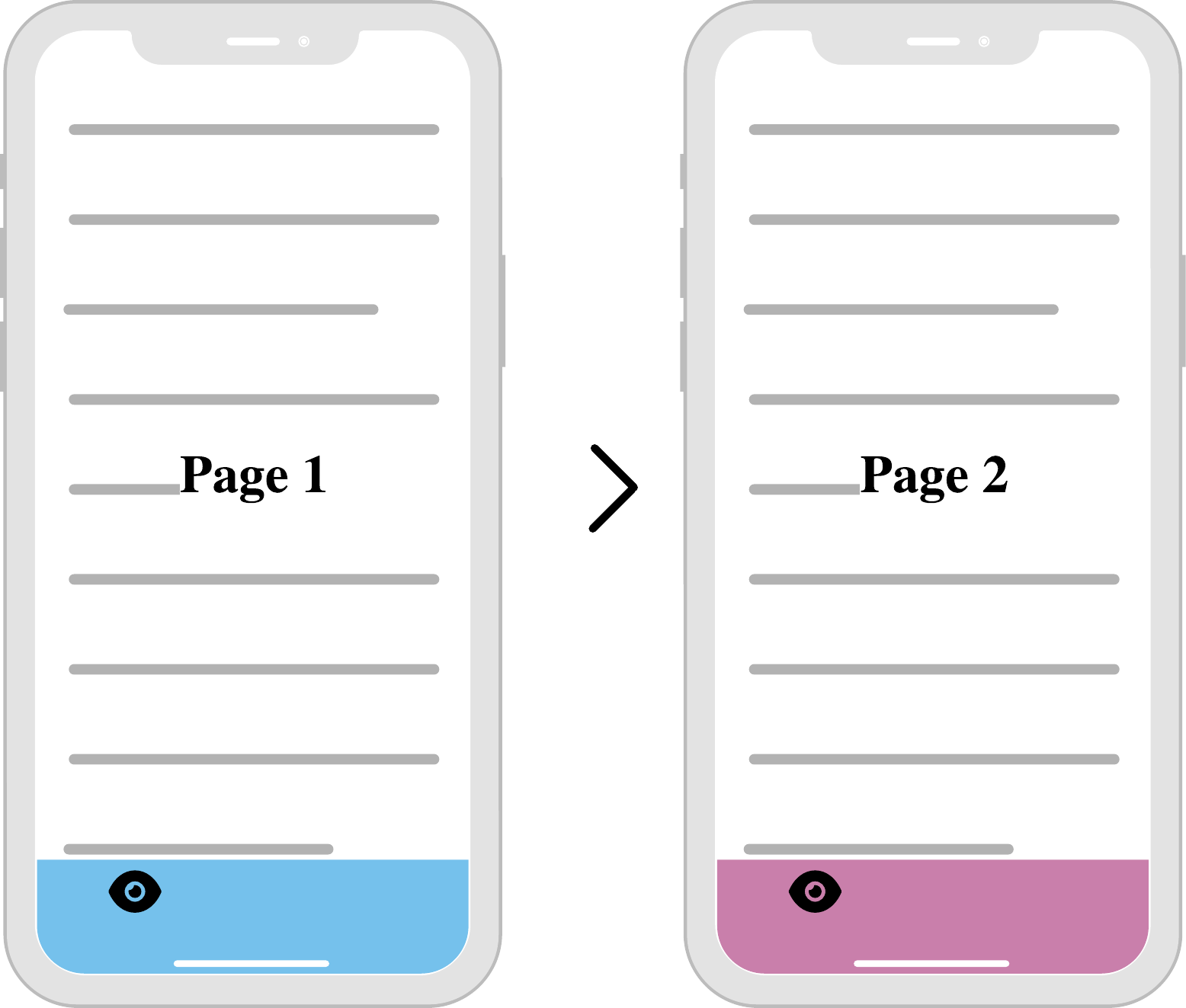}
\caption{Hitbox}
\label{fig:GazeB}
\end{subfigure}
\begin{subfigure}{0.3\textwidth}
\centering
\includegraphics[width=3.5cm, height=3cm]{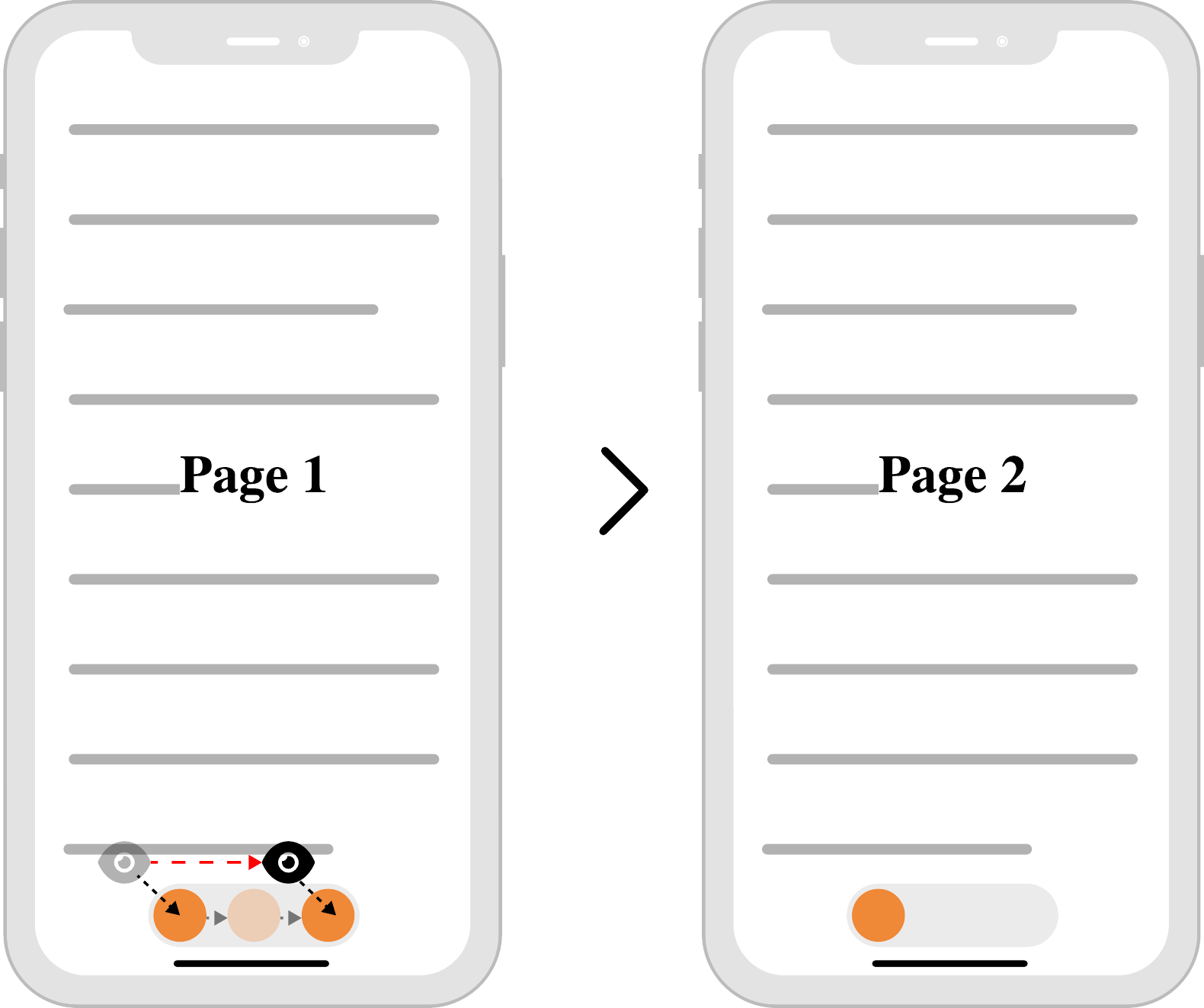}
\caption{Moving bar}
\label{fig:GazeD}
\end{subfigure}
\begin{subfigure}{0.3\textwidth}
\centering
\includegraphics[width=3.5cm, height=3cm]{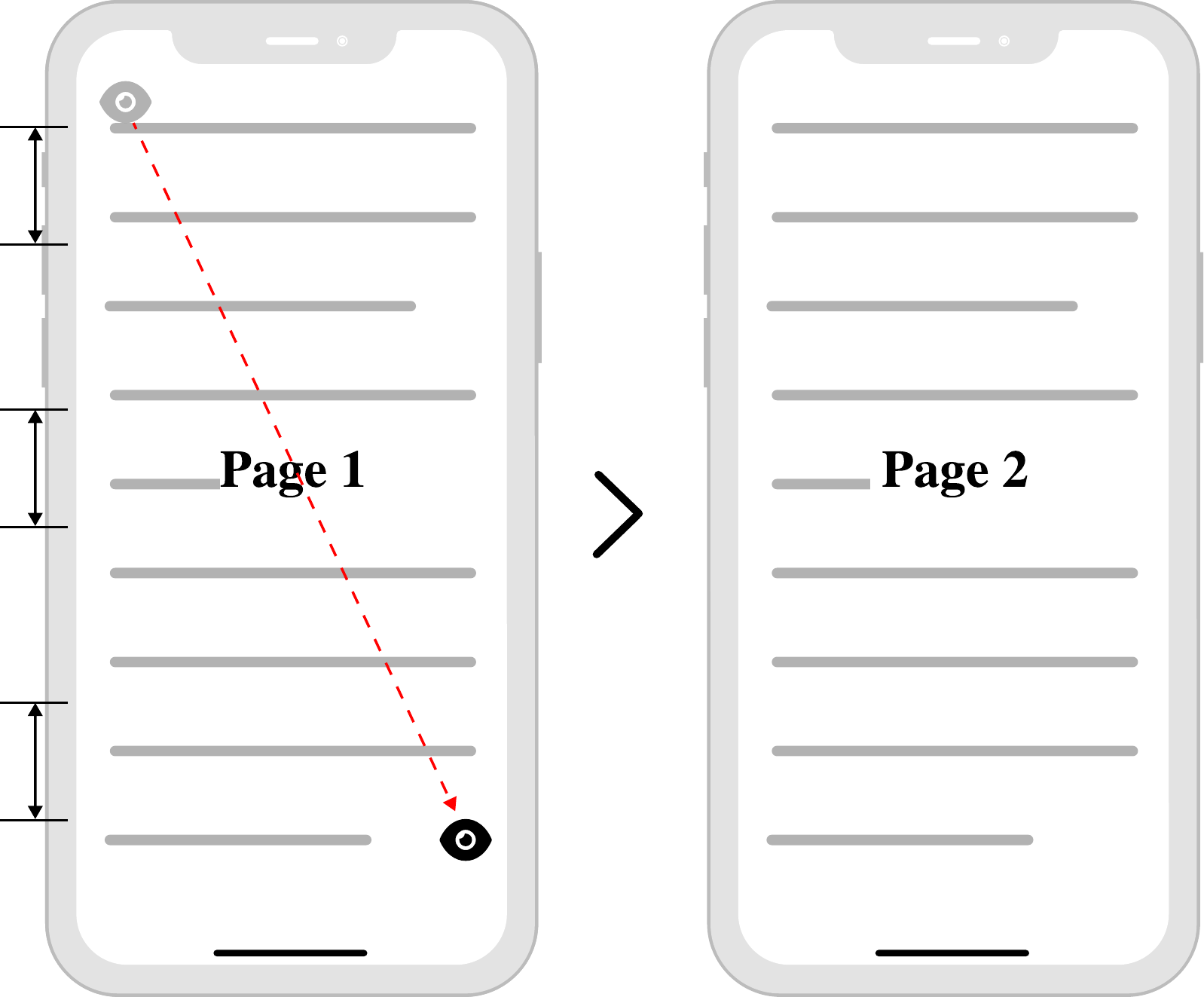}
\caption{Auto-Scrolling}
\label{fig:GazeA}
\end{subfigure}
\caption{(a) Hitbox: a Dwell-time-based explicit gaze scrolling technique where a reader is asked to fix their gaze at the bottom box for a period in order to trigger page scrolling action; (b) Moving bar: a Pursuit-based explicit gaze scrolling technique that requests readers to follow a moving bar to trigger page scrolling action; and (c) Auto-scrolling: an implicit gaze scrolling technique that predicts reading speed and turns the page automatically.}
\end{figure*}

\textit{Hitbox} is another explicit gaze scrolling technique based on dwell time; that is, when a reader finishes reading a page, they can fixate their gaze at the bottom box of the page, called Hitbox. During the gaze fixation, the colour of the box will change from light to dark. After a certain period of dwell time from 0.50s to 2.00s, the page will be automatically turned, shown in Fig~\ref{fig:GazeB}.  Shorter dwell times may come from random saccade and may afford faster operation manipulation, but the risk of errors due to false activation is higher, commonly recognised as the ``Midas Touch problem"~\cite{jacob1991use}. The period of dwell time threshold is set by the readers themselves and the recommended range is 0.72s to 1.50s~\cite{paivi2009fast, Mott2017improvingDwell}. To reduce accidental fixation, we exclude any gaze points whose dwell time is less than 300 milliseconds.

\textit{Moving bar} is an explicit pursuit-based gaze scrolling technique where a reader fixates their gaze on a bar and follows the bar to move from one end to the other at the bottom operation area, shown in Fig~\ref{fig:GazeD}. When the bar reaches the right end, the page will be turned. This involves two steps: activating the moving bar with a dwell time (300 ms) and then triggering page scrolling by successfully following the bar to the right end in a distance of 2.7cm within a period time. The period threshold is set by the readers themselves and the recommended range is 0.50s to 1.70s. To prevent accidental gaze falling on the bar area, only continuous points of gaze on the bar area are counted. Note that \textit{Hitbox} and \textit{Moving bar} interfaces are set at the bottom of the screen, which enables a natural transition from reading to actuating scrolling; however, the bottom is the least accurate region of the screen for gaze estimation as it is further away from the camera and it may only capture the eyelids of users. It would be interesting to assess the usability of gaze in these challenging regions.

\textit{Auto-Scrolling} is an implicit gaze scrolling technique that predicts the reading speed based on the observed gaze trajectory, shown in Fig~\ref{fig:GazeA}.
Specifically, we estimate the speed and time for a user to finish reading the last few lines based on the speed at which the user reads at the start and middle part of a page. Reading speed varies with individuals and materials, and we need to constantly monitor the gaze to be able to accurately predict their reading speed. However, constant gaze estimation can be costly in terms of computational power and battery, and lead to high latency. We adopt a trade-off approach that collects gaze points immediately after the page scrolling and samples gaze points after a few seconds. This design decision has taken two factors into account: the accuracy on the top region of the screen was often higher in our pilot test, as it is closer to the camera and has better coverage of face and eye images; and the reader's preferred reading region is in the middle of the page~\cite{sharmin2013reading}.

\subsection{Real-Time Gaze Estimation}
We adopt iTracker~\cite{krafka2016eyetracking} as our gaze estimation model, which is one of the most popular models for gaze estimation on mobile devices. Given an image from a camera, the system first crops the images for the face, left eye, and right eye, and then rescales them to the size of 224$\times$224 and 128 $\times$ 128 for face and eyes respectively. iTracker will take as input these rescale images and a face grid that is a 25 $\times$ 25 binary mask indicating the face position in the original image. 
The output of iTracker is the estimated gaze position as a coordinate, indicating their relative distance to the camera on a mobile device. We implement the iTracker model in PyTorch and train it on the GazeCapture dataset~\cite{krafka2016eyetracking}. The mean euclidean error of iTracker is 2.05cm on the test set of the GazeCapture dataset and 2.23cm on our test phone.

We use support vector regression (SVR) as a calibrator from GoogleGaze~\cite{valliappan2020accelerating}. We adopt the pursuit-based calibration method to balance time efficiency, using a moving dot to guide the user's gaze around the screen. More specifically, at the beginning of each experiment, we instruct a participant to fixate on one moving dot which crosses 125 points distributed around the boundary of a screen to maximise the mapping of the screen coordinate system. The image frames are processed and sent to iTracker to extract the features. The features along with the true point positions are fed to train a SVR model for calibration. We experiment with various calibration time durations, ranging from 50 frames (2 seconds) to 1500 frames (60 seconds), and find that fewer frames result in poorer performance, while more frames take longer to collect and train the model, offering only a marginal improvement in accuracy. In the end, we opt for a 5-second calibration process, collecting 125 frames. The pilot test on our device has shown a mean Euclidean error of 0.95cm for sitting and 1.98cm for walking after calibration.

\subsection{Implementation and System Deployment}\label{subsec:deployment}
\begin{figure}[!htbp]
    \centering
    \includegraphics[width=0.55\textwidth]{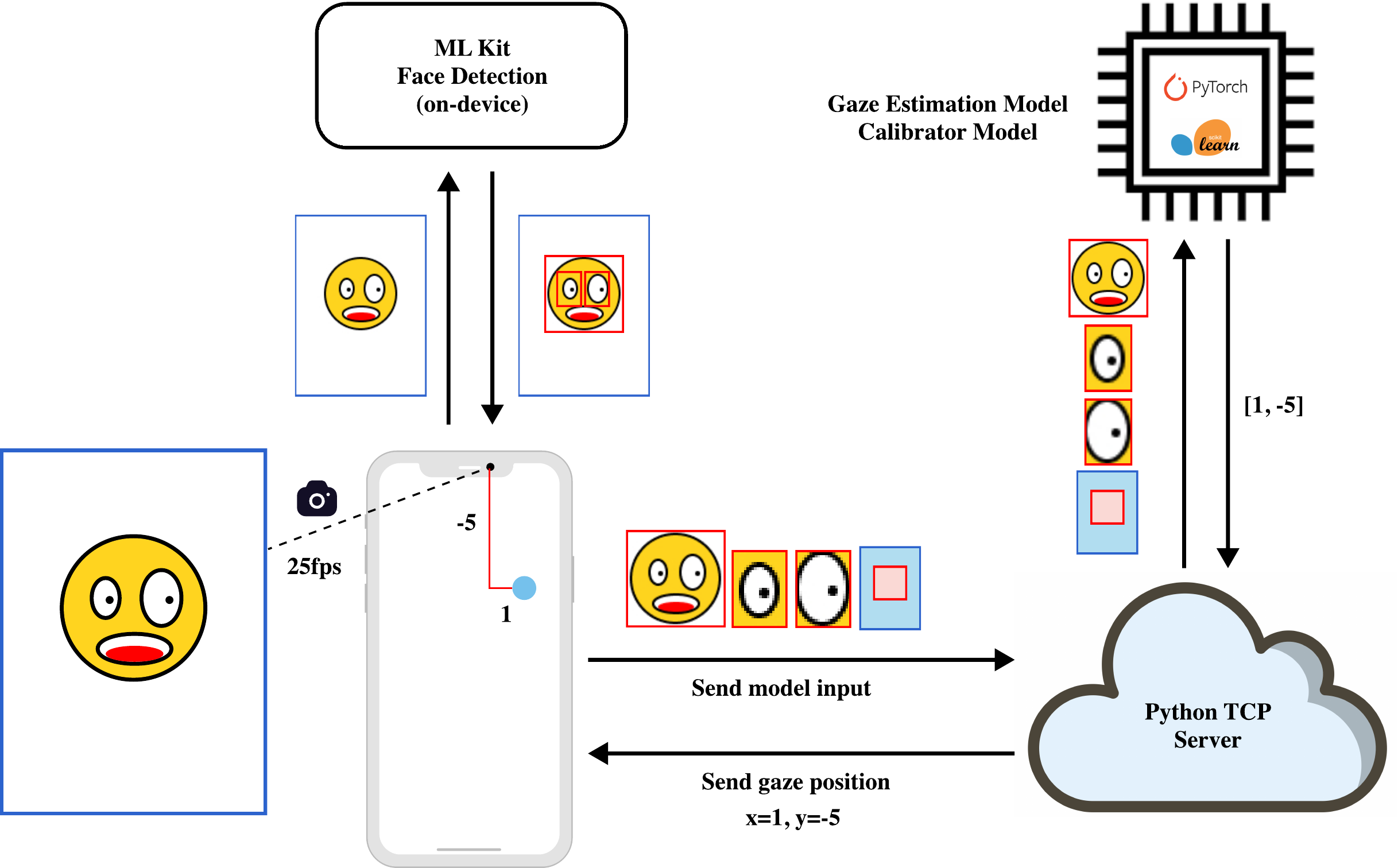}
    \caption{System Architecture}
    \label{fig:system}
\end{figure}

We implement our Gaze Reading application in Flutter~\cite{Google2022flutter}, a cross-platform mobile app development framework built by Google. To facilitate debugging and testing, the system consists of an app deployed on the iOS device as client-side and an inference service on the server as server-side, both connected via a local area network (LAN), as shown in Fig~\ref{fig:system}.

First, images are captured of the user from the device’s front-facing camera; that is, 25 frames per second. For each image, we use ML Kit~\cite{Google2022mlkit} to detect face and eyes, which will be used to create the input required for the relevant gaze estimation model. ML Kit is a machine learning toolkit for iOS and Android~\cite{Google2022mlkit} and it provides facial landmark and contour detection. With ML Kit, we process each input image and generate the face and eye images and face grid. The total processing time is between 10ms and 25ms. 

The server that runs the iTracker model and SVR model receives the input and infers with the gaze position. After processing, all the inputs will be deleted from the server and the inferred gaze points along with timestamps are stored for further analysis. The data transport latency ranges around 7ms-50ms and the model inference latency is 60-75ms, which is the highest in the whole system.

\section{User Study Design}\label{sec:userstudy}
We design our user study to answer the following questions: 
\begin{itemize}
    \item \textbf{Q1}: Which gaze scrolling technique is most preferred by users in a reading application? 
    \item \textbf{Q2}: How do gaze scrolling techniques perform in a mobile condition and which technique performs the best? 
    \item \textbf{Q3}: Is there any learning effect of using gaze scrolling techniques? 
\end{itemize}

To address these questions, we evaluate the differences in user experience between touch and gaze-based scrolling and investigate the impact of mobility on the effectiveness of gaze interfaces. We design and conduct a within-subjects user study where we ask each participant to use each of these five scrolling techniques. We use a $4 \times 4$ Latin Square to balance the sequential effect between the four gaze scrolling techniques.

\subsection{Participant}
We recruit 20 participants (13 male and 7 female) from the undergraduate and postgraduate population at our university. The age of these participants ranges from 20 to 31 years, with a mean of 25.05 and a standard deviation of 6. They all have normal eyesight or vision correction and 14 of 20 wear glasses,  with left eye myopia ($M$ = 399.4, $SD$ = 185.2) and right eye myopia ($M$ = 424.5, $SD$ = 181.9). From the demographic questionnaire, English language proficiency is assessed on a scale ranging from 1 as Elementary proficiency to 5 as native or bilingual level. The participants' English level is assessed with mean of 3.2 \hl{($SD$ = 1.05)}. Their familiarity with gaze technologies has the mean of 1.5 \hl{($SD$ = 0.5)} within the range of 1 and 5. 

The study is approved by the university ethics committee under the reference number CS-15883. We provide each participant with an information sheet stating the objective of our user study and what data will be collected. They have 24 hours to consider whether to take part. Each experiment takes approximately 1 to 1.5 hours and each participant is compensated with a £10 Amazon voucher.

\subsection{Text Selection}
We select 10 text samples from 5 categories including Fiction, Nonfiction, Fairy Tale, Fantasy Story and Poetry. Each sample contains around 5000 words and has 5 or 6 pages in the reading application. The content collection is evaluated by Lexile Index~\cite{stenner1996measuring}($M$ = 1065L, $SD$= 265). The higher the index, the more difficult a reading sample is. The selection caters for a wide range of leisure reading materials to simulate different reading scenarios on handheld mobile devices.

\subsection{Apparatus}
We conduct our user study on an iPhone 13 Pro Max (6.7-inch, 240 grams, 256GB, iOS 16.0.2), with a display of 428 pixels in width and 926 pixels in height. We left the top area (i.e., 100 pixel in height) for instructions and activation signals, and the bottom area (i.e., 150 pixels in height) for gaze interfaces. The remaining area (i.e., 676 pixels in height) is for reading. All versions of the reading interface use 17 font size, 2-line space, and black font colour. 

\subsection{Experimental Procedure}
Each participant was asked to use touch and four gaze-based scrolling techniques: \textit{Eye-Swipe}, \textit{Hitbox}, \textit{Moving bar}, and \textit{Auto-Scrolling} in a sequence under both sitting and walking conditions. For each sequence, the order of the gaze techniques was counterbalanced while touch-based scrolling was always the first to allow the participants to familiarise themselves with the reading application. Touch-based scrolling also served as the baseline condition for training and we do not expect significant learning effects or bias from this as most participants likely have used touch interfaces already. This touch interface has two buttons at the bottom of the screen on the left and right edges with a distance of 360 pixels between them, for turning the page backwards and forwards respectively. Participants were advised to simulate a single-hand scenario: using one hand to hold the phone and scroll the page. However, some participants found the phone too big or too heavy, and they opted to use one hand to hold and the other hand to scroll.

At the end of each scrolling technique session, each participant is asked to fill in the questionnaires of System Usability Scale (SUS) and NASA-Task Load Index (NASA-TLX). SUS contains 10 usability questions, each being assessed on a scale from 1 (strongly disagree) to 7 (strongly agree). The higher the score, the better the perceived usability. We also use 7-likert NASA-TLX to assess the task load on 6 dimensions including mental demand, physical demand, temporal demand, performance, effort, and frustration. Participants self-rate each dimension on a scale from 1 (lowest load) to 7 (highest load).

At the end of an experiment, a semi-structured interview is conducted. SUS is a standardised questionnaire for assessment of perceived usability and has been widely used in HCI for user interface evaluation and has proven its validity~\cite{brooke1996sus, Lewis2018sus}. We use a standard version of SUS~\cite{wilson2018autopager} and add the following 6 questions for \hl{collecting subjective experience of mobility conditions (sitting and walking),} in Table~\ref{tab:q1116}. NASA-TLX~\cite{hart1986nasa} was originally used to evaluate subjective workloads for simulations and flight experiments, and has also been used extensively to evaluate user workload requirements in user interfaces~\cite{Hart2006NASATLX20, Yuan2013Evaluation}. In a semi-structured questionnaire, we ask each participant to rank their preference and record their experience on all the five scrolling techniques.

\begin{table*}[!htbp]
\caption{Additional questions of SUS for Subjective Experience of Mobility Conditions} 
\label{tab:q1116}
\begin{tabular}{|l|l|}
\hline
\multicolumn{1}{|c|}{No.} & \multicolumn{1}{c|}{Questions} \\ \hline
Q11/14 & I was aware of when new text is introduced to the display when sitting / walking   \\ \hline
Q12/15 & I could read as slow as I like when sitting / walking  \\ \hline
Q13/16 & I felt in control while reading when sitting / walking\\ \hline
\end{tabular}

\end{table*}

The experiments are conducted in a lab and the participants are asked to hold the phone to read in their most natural, comfortable postures. For the sitting condition, they can either sit up or rest on the chair. For the walking condition, they are asked to walk at their natural speed along the marked space in the lab. More details of the experiment formulation has been provided in the supplementary materials.

\section{Results}\label{sec:results}
This section will analyse the results to answer the questions in Section~\ref{sec:userstudy}. We start from overall comparison between touch and gaze scrolling techniques, investigate the effect of mobility conditions on the use of scrolling techniques, study the learning effect on gaze scrolling techniques, and look into gaze patterns using scan-path and heatmaps. 

\subsection{Analytical Plan}
This section will describe the analytical plan based on the research questions in Section~\ref{sec:userstudy}, including hypotheses and analysis methodologies. 

For \textbf{Q1} on comparison of different scrolling techniques, we hypothesise that there is significant difference in usability, cognitive load, and reading behaviours between these scrolling techniques. To test, we run one-way repeated measures ANOVA on SUS, NASA-TLX (the overall scores), and the reading time with the scrolling techniques as the independent variable. Also we analyse the semi-structured interview for qualitative insight. The results are reported in Section~\ref{subsection:comparison_of_touch_gaze}.

For \textbf{Q2} on the impact of mobility, we hypothesise that there is significant difference in the usability of scrolling techniques between the sitting and walking conditions. We run two-way repeated measures ANOVA with scrolling techniques and mobility conditions as the independent variables on the scores from additional Likert questions of SUS for Subjective Experience of Mobility Conditions in Table~\ref{tab:q1116}, and on the reading time under different mobility conditions. We then analyse the semi-structured interview data and compare their gaze scan-path and heatmaps. The results are reported in Section~\ref{subsec:impact_mobility} and \ref{subsec:gaze_path}.

For \textbf{Q3} on the learning effect, we hypothesise that there is a learning effect among the gaze scrolling techniques; that is, after experimenting one type of gaze interface, participants can become more proficient at using the other type of gaze interfaces. To test, we run two-way repeated measures ANOVA on SUS and NASA-TLX (the overall scores) with scrolling techniques and experiment order of gaze techniques as the independent variables, and three-way repeated measures ANOVA  on reading time with scrolling techniques, mobility conditions and experiment order of gaze techniques as the independent variables. The results are reported in Section~\ref{subsec:learnign_effect}.

\subsection{Comparison of Touch and Gaze Scrolling Techniques }\label{subsection:comparison_of_touch_gaze}
\begin{figure}[!htbp]
    \centering
    \begin{subfigure}{0.386\textwidth}
    \centering
    \includegraphics[width=\textwidth]{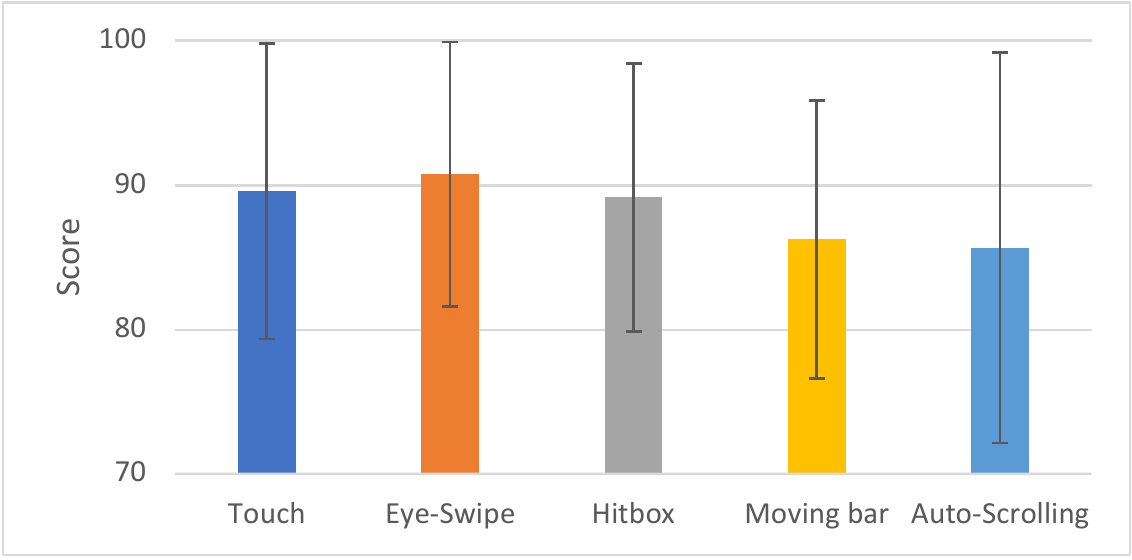}
    \caption{SUS of scrolling techniques}
    \label{fig:sustype}
    \end{subfigure}
    \begin{subfigure}{0.389\textwidth}
    \centering
    \includegraphics[width=\textwidth]{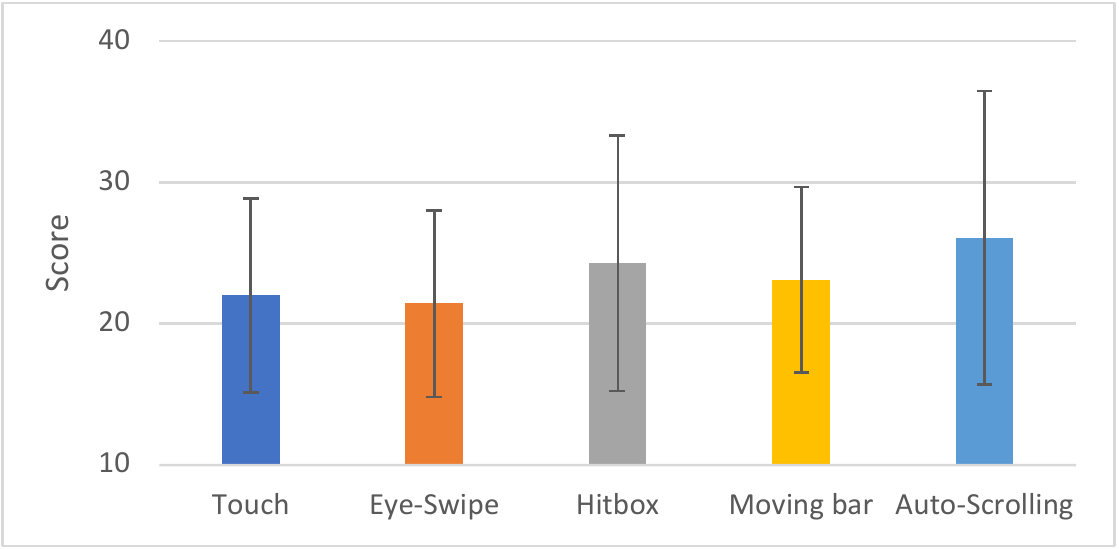}
    \caption{NASA-TLX of scrolling techniques}
    \label{fig:nasatype}
    \end{subfigure}
    \caption{Overall comparison of scrolling techniques}
    \label{fig:overall}
\end{figure}

To assess the usability of gaze-assisted scrolling techniques, we look into the following measures. First, on the SUS result, \textit{Eye-Swipe} gets the highest score ($M$= 90.75, $SD$= 9.15), and \textit{Auto-Scrolling} has the lowest score ($M$= 85.65, $SD$= 13.50). For the other scrolling techniques, \textit{Touch} has the SUS score ($M$= 89.55, $SD$= 10.25), \textit{Hitbox} ($M$= 89.20, $SD$= 9.30), and \textit{Moving bar} ($M$= 86.25, $SD$= 9.60). The comparison of SUS scores are presented in Fig~\ref{fig:sustype}.

Secondly, on the NASA-TLX scale, \textit{Eye-Swipe} has the lowest score ($M$= 21.43, $SD$= 6.60), and \textit{Auto-Scrolling} has the highest score ($M$= 26.97, $SD$= 10.41). For the other scrolling techniques,  \textit{Touch} ($M$= 22.02, $SD$= 6.86), \textit{Hitbox} ($M$= 24.29, $SD$= 9.03), \textit{Moving bar} ($M$= 23.10, $SD$= 6.56). The comparison of NASA-TLX scores is presented in Fig~\ref{fig:nasatype}. The one-way repeated measures ANOVA shows that there is no significant difference between these five scrolling techniques on  NASA-TLX  ($\it{F}(4, 76) = 1.913, \it{p} = .117$) and  SUS ($\it{F}(4, 76) = 2.453, \it{p} = .053$).

\begin{figure}[!htbp]
    \begin{subfigure}{0.385\textwidth}
        \centering
    \includegraphics[width=\textwidth, height = 90pt]{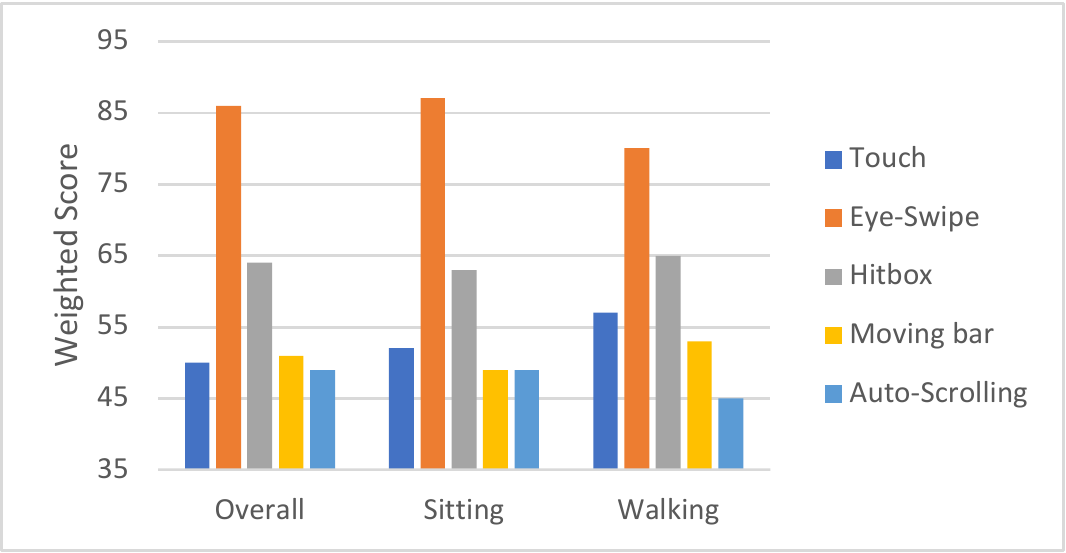}
    \caption{Subjective Ranking}
    \label{fig:ranks}
    \end{subfigure}
    \begin{subfigure}{0.385\textwidth}
        \centering
         \includegraphics[width=\textwidth, height = 90pt]{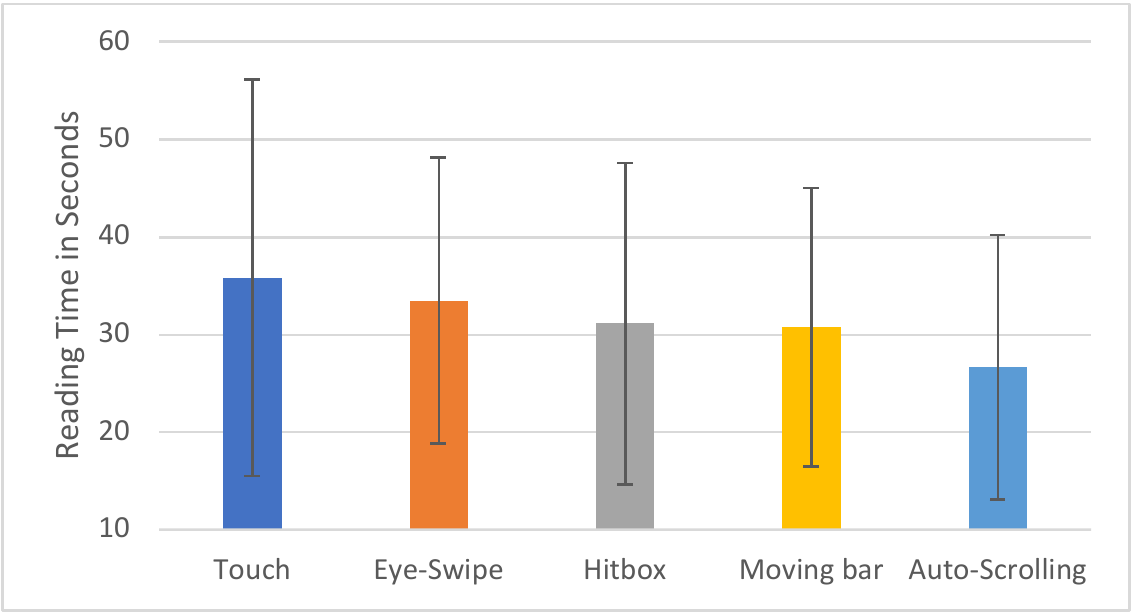}
    \caption{Reading Time of Per Page}
    \label{fig:pageturntype}
    \end{subfigure}
    \caption{Compare User Subjective Ranking on their preferred scrolling techniques and reading time}
    \label{fig:preferred_scrolling}
\end{figure}

In the semi-structured questionnaire, \textit{Eye-Swipe} is ranked highest for being the most preferred scrolling technique: 11 times (55\%) being ranked first. Figure~\ref{fig:ranks} compares the weighted ranks of the scrolling techniques, where we assign a score 5 for being ranked first, 4 for second, and 1 for the last. As we can see, \textit{Eye-Swipe} has the highest score, and \textit{Eye-Swipe} and \textit{Hitbox} are consistently better perceived than \textit{Touch} under the different mobility conditions. Participants have commented \textit{Eye-Swipe} as ``a fun and easy way to read''. 



In addition, we compare average reading time per page (RTPP) between these scrolling techniques in Fig~\ref{fig:pageturntype}. Individuals can have different reading speeds on different materials because of their own reading patterns, such as pauses, regressions (re-read), slowdowns and line skips. However, we hypothesise that the modality of interfaces (Touch vs Gaze) will not impact users' reading patterns; that is, they can read naturally in their own pace, no matter which scrolling technique is adopted. We conduct a one-way repeated measure ANOVA of overall mean of RTPP with scrolling techniques as the independent variable, while not differentiating between walking and sitting. There is a significant difference in RTPP among the five scrolling techniques ($\it{F}(2.55, 48.36) = 4.616, \it{p} = .009$). The post hoc $\it{t}$-test with Bonferroni correction demonstrates that \textit{Auto-scrolling} has significantly shorter reading time compared to the \textit{Touch} method ($\it{t(39)} = -4.07, \it{p} = .002$), and there are no significant difference between the other scrolling techniques.

Except for \textit{Auto-Scrolling}, the other gaze interfaces have similar reading time to \textit{Touch}. More specifically, \textit{Touch} has the reading time ($M$= 35.81s, $SD$= 20.30), followed by \textit{Eye-Swipe} ($M$= 33.50s, $SD$= 14.66), \textit{Hitbox} ($M$= 31.15s, $SD$= 16.47), \textit{Moving bar} ($M$= 30.75s, $SD$= 14.30), and \textit{Auto-Scrolling} ($M$= 26.65s, $SD$= 13.54). \textit{Auto-Scrolling} has the shortest page-turning time and causes frustration for users; that is, the page is turned sometimes before a user finishes. This is reflected in its low SUS score and high NASA-TLX score. 

\subsection{Impact of Mobility on Gaze}\label{subsec:impact_mobility}
We compare the above measures under the sitting and walking conditions to find out which gaze interface is most significantly impacted by the mobility. We will first compare the scrolling techniques under each condition and then between conditions.

Following Table~\ref{tab:q1116}, we group the questions as (11, 14), (12, 15), and (13, 16), each pair corresponding to the same question under the sitting and walking conditions. We run two-way repeated measures ANOVA with scrolling techniques and mobility conditions as the two independent variables. The results showed that subjective experience has a significant difference among the five scrolling techniques ($\it{F}(4, 236) = 15.75, \it{p} < .001 $), and a significant difference on mobility conditions ($\it{F}(1, 59) = 57.302, \it{p} < .001 $), but there is no significant difference in their interactions ($\it{F}(4, 236) = 2.272, \it{p} = .062 $).

For post hoc comparison, we run the pairwise Wilcoxon test instead of $\it{t}$-test, because the Likert score is ordinal rather than continuous. The results suggest that there is significant difference between sitting and walking conditions ($\it{W} = 56257.5, \it{p} < .001 $) and there is significant difference between \textit{Auto-Scrolling} and \textit{Hitbox} ($\it{W} = 5256, \it{p} < .001 $), \textit{Auto-Scrolling} and \textit{Eye-Swipe} ($\it{W} = 4564, \it{p} < .001 $), \textit{Auto-Scrolling} and \textit{Moving bar} ($\it{W} = 5244, \it{p} < .001 $), \textit{Auto-Scrolling} and \textit{Touch} ($\it{W} = 4877, \it{p} < .001 $), while there is no significant difference between the remaining scrolling techniques.

We compare the user ranking of the scrolling techniques under the sitting and walking conditions in Figure~\ref{fig:ranks}. From sitting to walking, the rank of \textit{Touch}, \textit{Hitbox}, and \textit{Moving bar} increases while the rank of \textit{Eye-Swipe} and \textit{Auto-Scrolling} decreases. However, the overall ranking of preferred scrolling techniques is still the same. It suggests that \textit{Eye-Swipe} and \textit{Hitbox} can still perform well under the walking condition, more robust and tolerant to unstable gaze estimation. 

\begin{figure}[!htbp]
    \centering
    \includegraphics[width=0.385\textwidth]{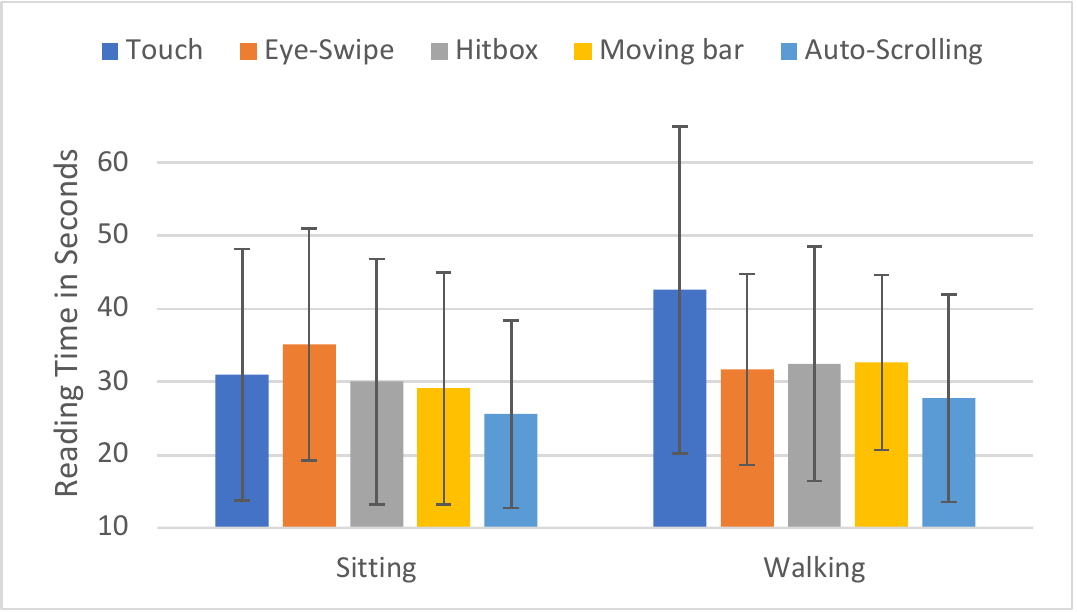}
    \caption{Comparison of reading time per page under sitting and walking conditions}
    \label{fig:pageturnposture}
\end{figure}

We conduct two-way repeated measure ANOVA of RTPP with scrolling techniques and mobility conditions being the two independent variables. The result shows that the mobility condition itself ($\it{F}(1, 19) = 3.894, \it{p} = .063$) does not affect the reading time, but the interactions between scrolling techniques and mobility conditions ($\it{F}(4, 76) = 3.844, \it{p} = .007$) have a significant effect on RTPP. Under the walking condition, \textit{Auto-scrolling} also has significantly shorter reading time compared to \textit{Touch} ($\it{t(19)} = -3.53, \it{p} = .022$), while no significant differences under the sitting condition. When using \textit{Touch} scrolling technique, we find that the RTPP of walking is significantly longer than when sitting ($\textit{t(19)} = -4.23, \textit{p} < .001$), with no significant differences observed across the gaze scrolling techniques.

\subsection{Gaze Trajectories and Patterns}\label{subsec:gaze_path}

\begin{figure*}[!htbp]
    \centering
    \includegraphics[width=0.93\textwidth]{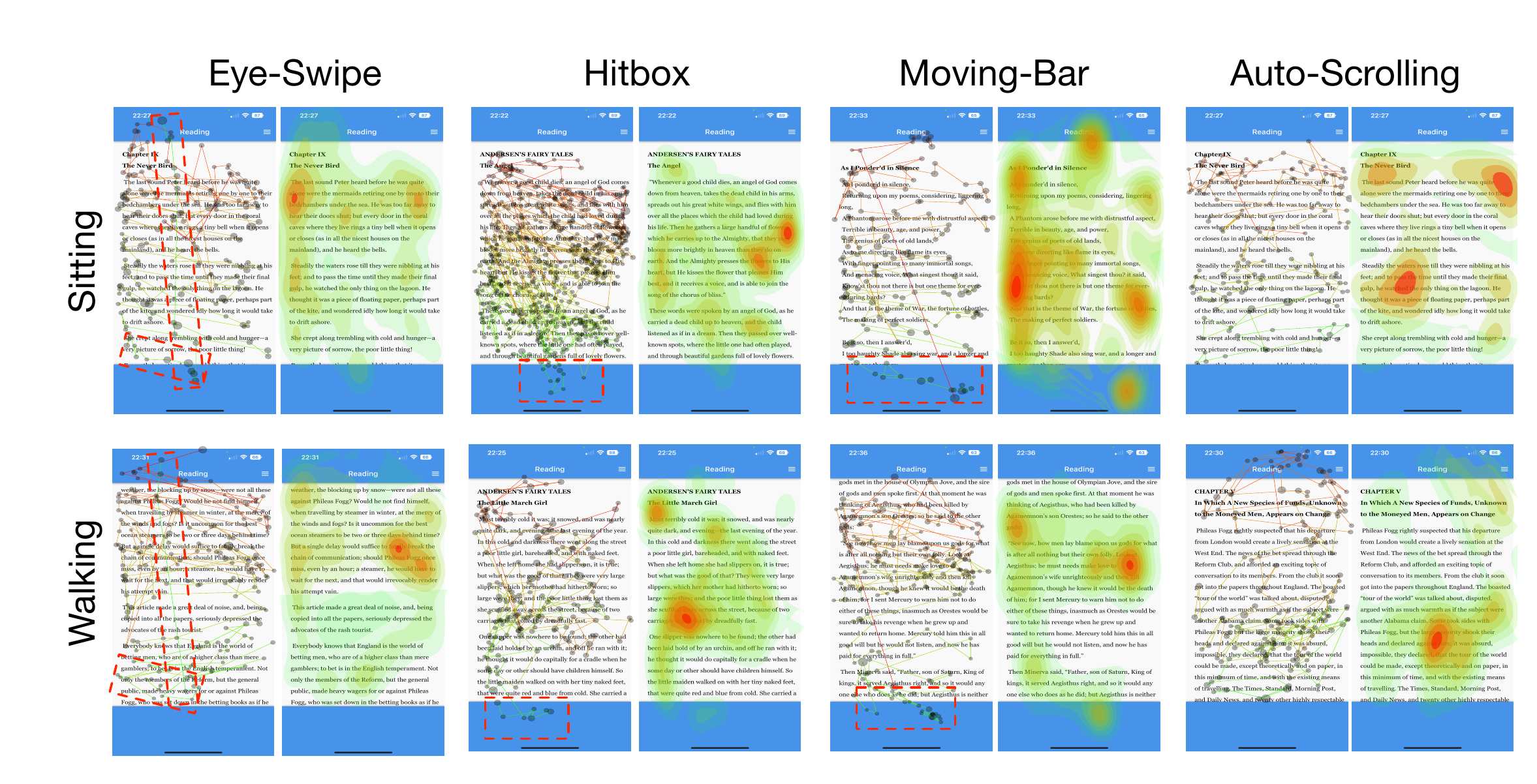}
    \caption{Gaze scan-path and heatmap of gaze scrolling techniques under sitting and walking. A scan-path image shows a trajectory of gaze points on a page and its colour changing from red to green indicates the transition. A heatmap image visualises the clusters of gaze points. For the explicit gaze interaction methods, we mark the gaze patterns that activate the scrolling technique on the scan-path images.}
    \label{fig:gaze-pattern}
\end{figure*}

We present gaze scan-path and heatmaps on these gaze interfaces under both sitting and walking conditions in Figure~\ref{fig:gaze-pattern}. First of all, we can observe gaze estimation works well in that gaze covers all the text and 
gaze patterns that activate the explicit gaze interfaces are visible. Especially, \textit{Eye-Swipe} -- a green vertical trajectory from the bottom to the page, and \textit{Hitbox} -- a cluster of consecutive gaze points at the bottom area, are robust to the inaccuracy of gaze estimation. The results are really promising, demonstrating the potential of the use of gaze estimation in real-time. 

\begin{figure*}
    \centering
    \begin{subfigure}{0.23\textwidth}
\centering
    \includegraphics[width=50pt, height=100pt]{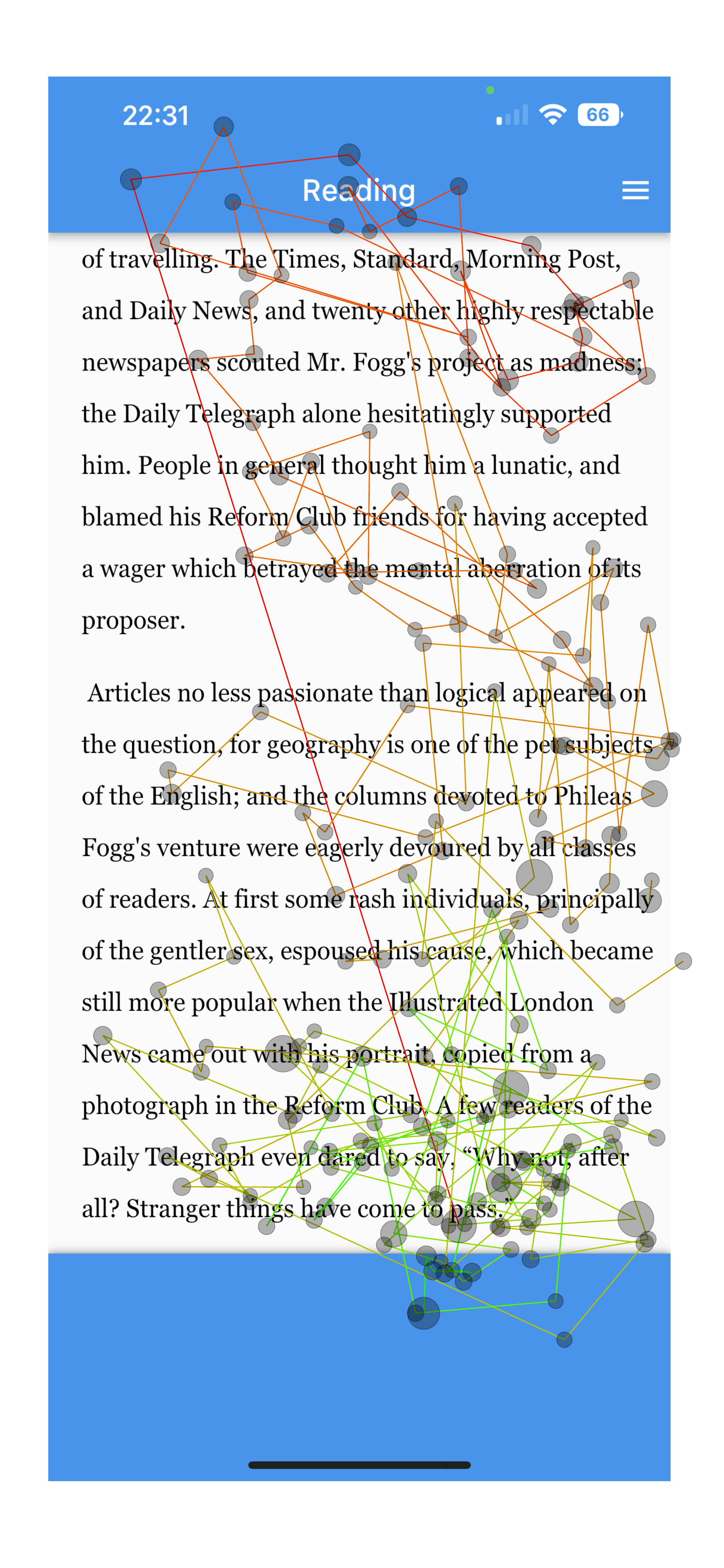}
    \caption{Scan-path \textit{Moving-bar}}
    \label{fig:Scan-GBP1-D-Walking-2}
    \end{subfigure}
        \begin{subfigure}{0.23\textwidth}
  \centering
    \includegraphics[width=50pt, height=100pt]{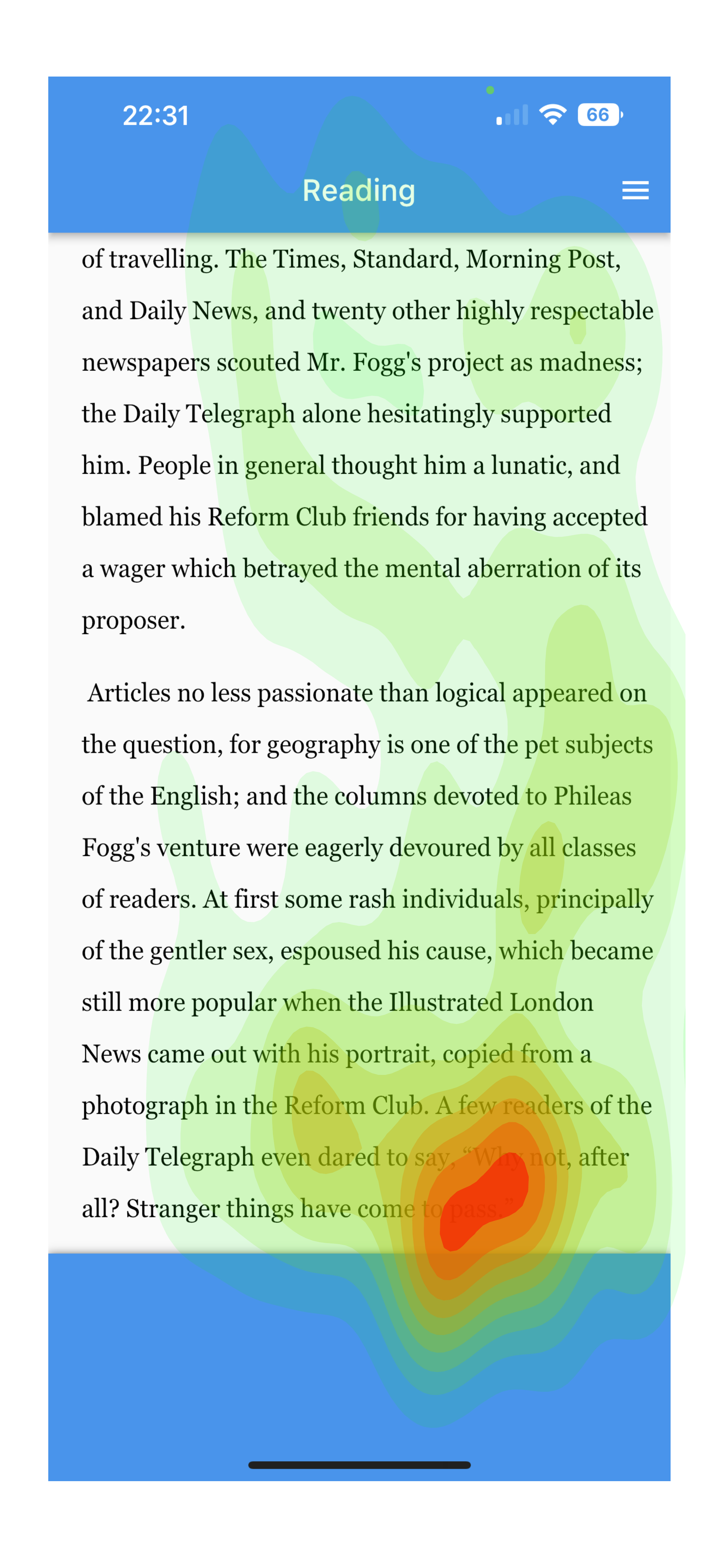}
    \caption{Heatmap \textit{Moving bar}}
    \label{fig:Heat-GBP1-D-Walking-2}
    \end{subfigure}
        \begin{subfigure}{0.23\textwidth}
   \centering
    \includegraphics[width=50pt, height=100pt]{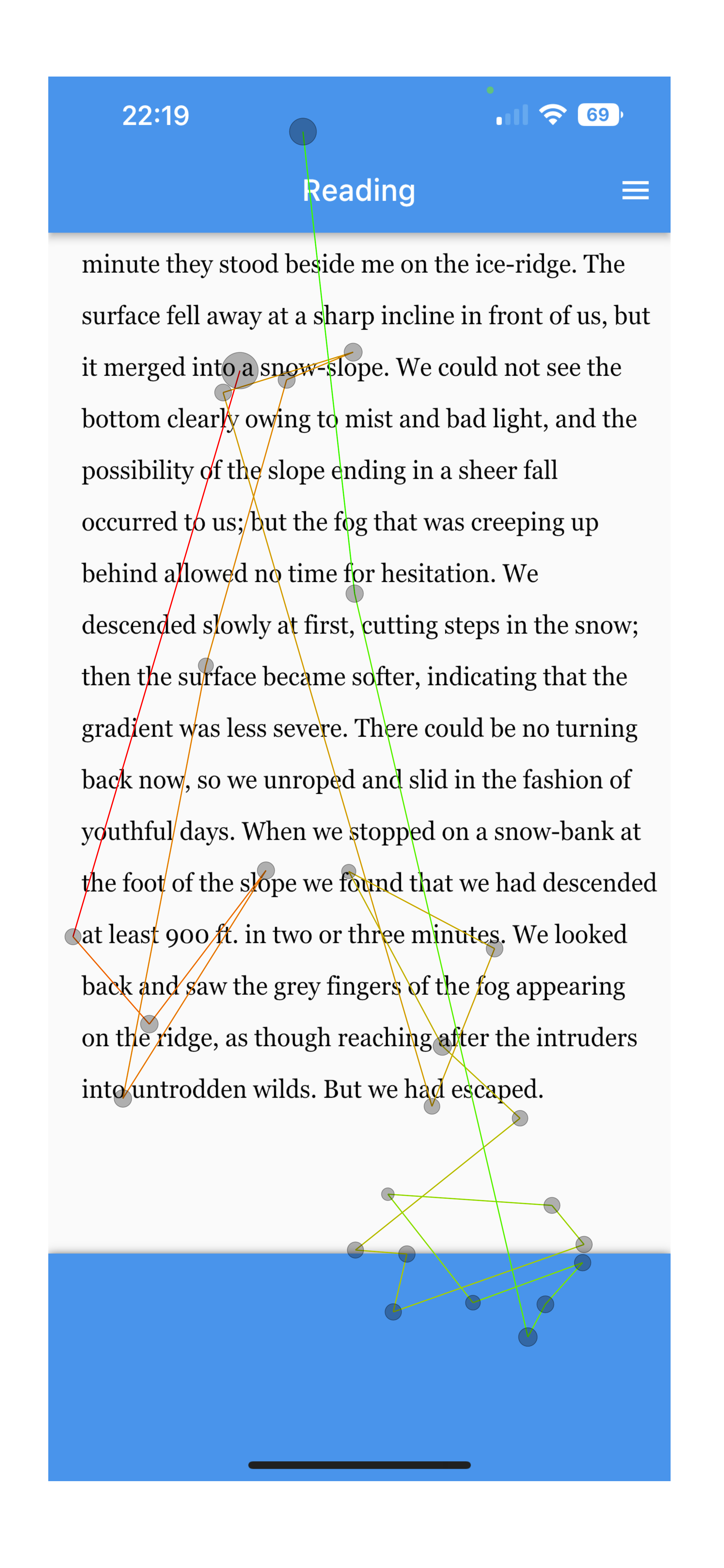}
    \caption{Scan-path \textit{Eye-Swipe}}
    \label{fig:Scan-GCP3-C-walking-6}
    \end{subfigure}
        \begin{subfigure}{0.23\textwidth}
   \centering
    \includegraphics[width=50pt, height=100pt]{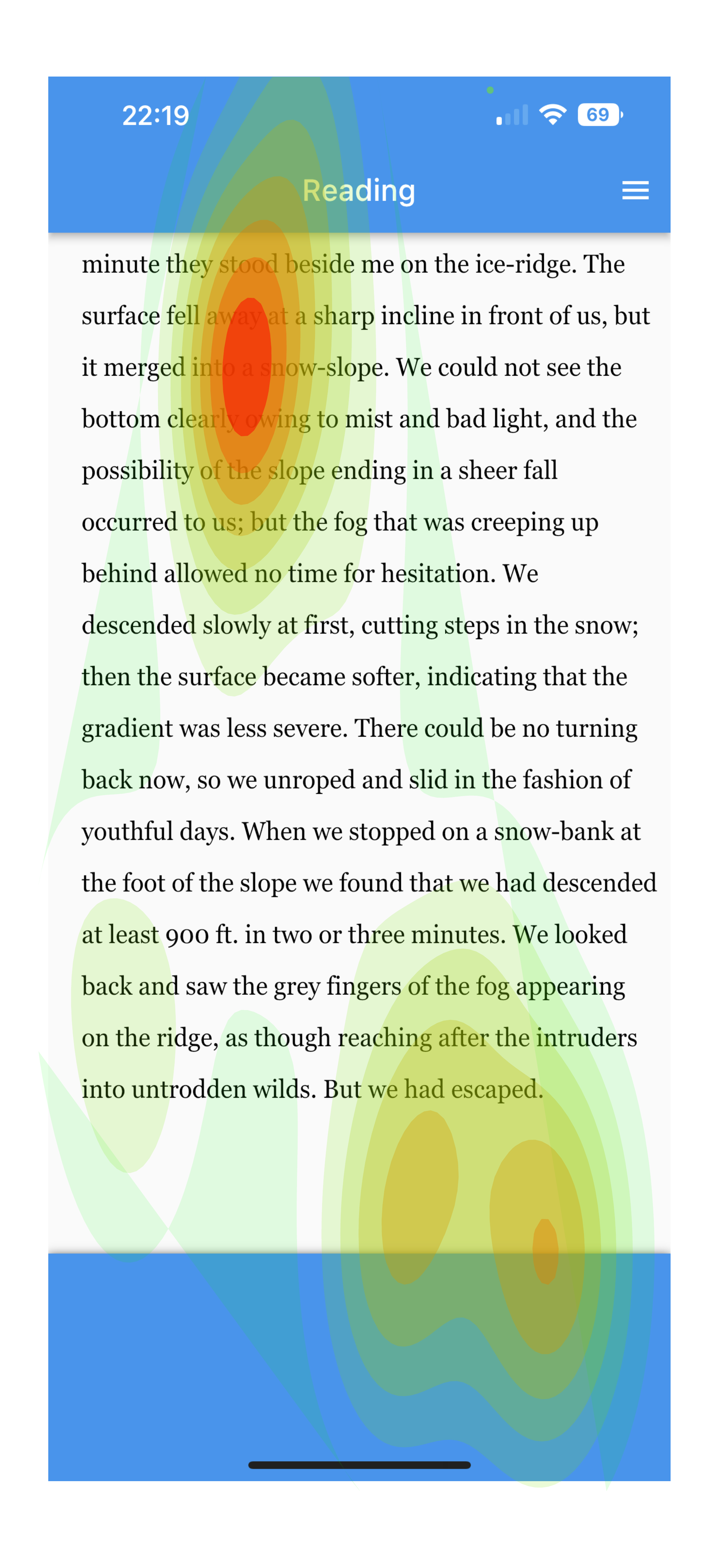}
    \caption{Heatmap \textit{Eye-Swipe}}
    \label{fig:Heat-GCP3-C-walking-6}
    \end{subfigure}
    \caption{Challenging examples of gaze patterns observed under walking condition}
    \label{fig:bad-pattern}
\end{figure*}

Secondly, under the walking condition, gaze patterns are more sparse and more likely out of the reading area. This makes it challenging to predict the correct reading speed for \textit{Auto-Scrolling}, the implicit gaze interface. \textit{Moving bar} is difficult to operate as it can abort once the gaze jumps out of the area. Participants feel frustrated trying to get it working several times. Thirdly, gaze patterns can be skewed to one side during walking, as sometimes participants tilt the phone, which does not capture the right, front face for gaze estimation. We have shared a few challenging examples of gaze patterns in Figure~\ref{fig:bad-pattern}. 

\subsection{Learning Effect of Gaze Interfaces}\label{subsec:learnign_effect}

\begin{figure*}[!htbp]
    \centering
    \begin{subfigure}{0.32\textwidth}
    \includegraphics[width=\textwidth]{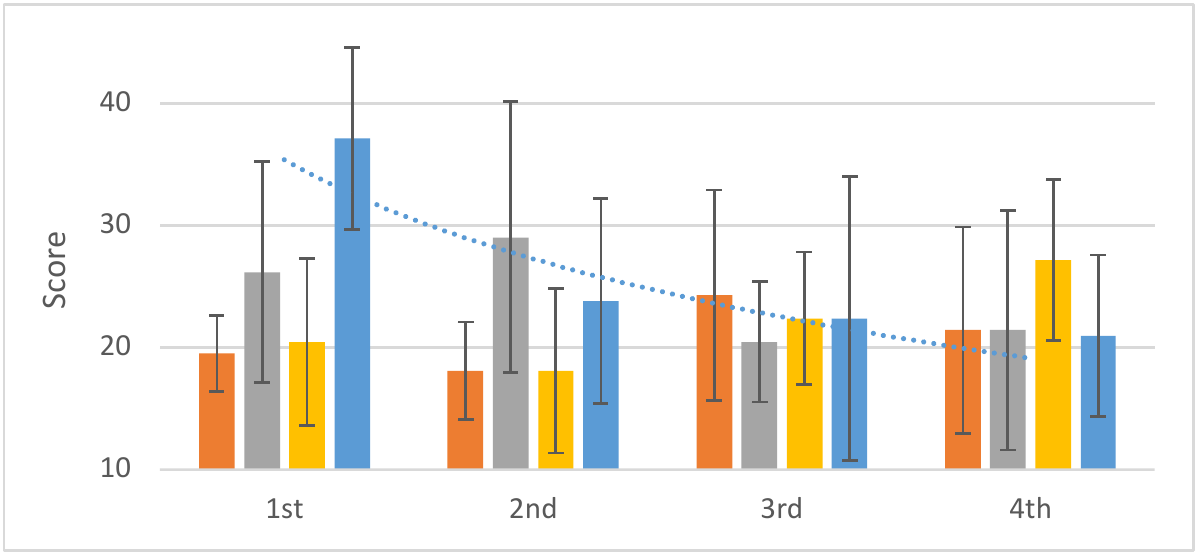}
    \caption{NASA-TLX}
    \label{fig:seqnasa}
    \end{subfigure}
    \begin{subfigure}{0.331\textwidth}
    \includegraphics[width=\textwidth]{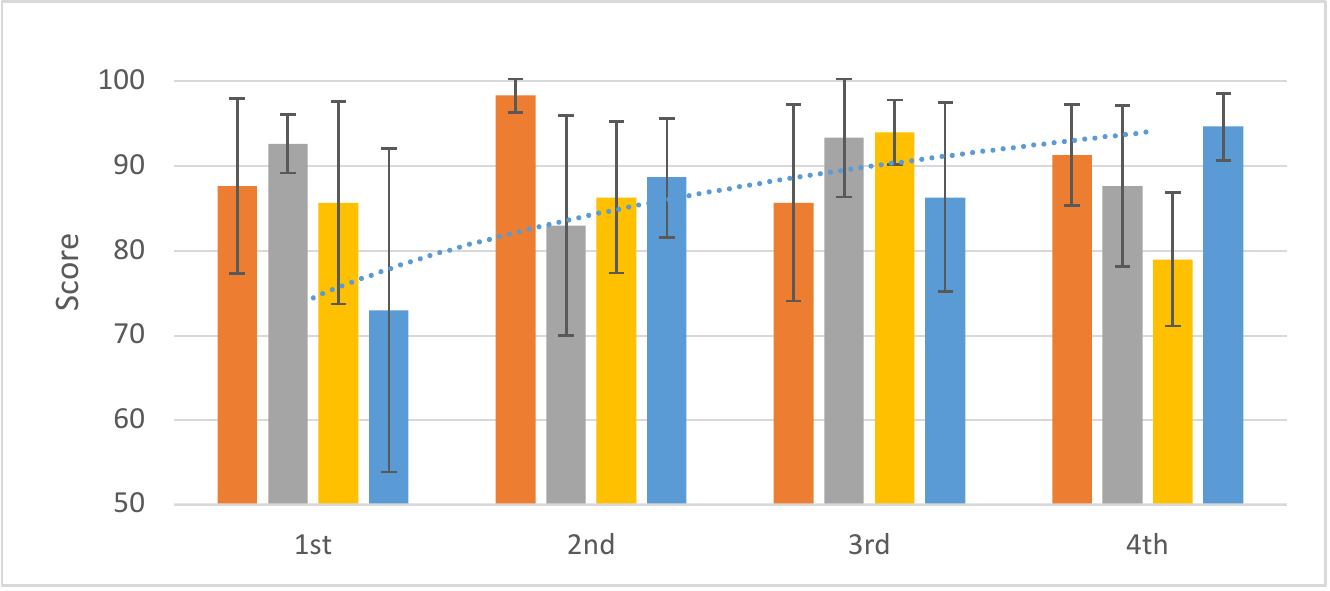}
    \caption{SUS}
    \label{fig:seqsus}
    \end{subfigure}
    \begin{subfigure}{0.32\textwidth}
    \centering
    \includegraphics[width=\textwidth]{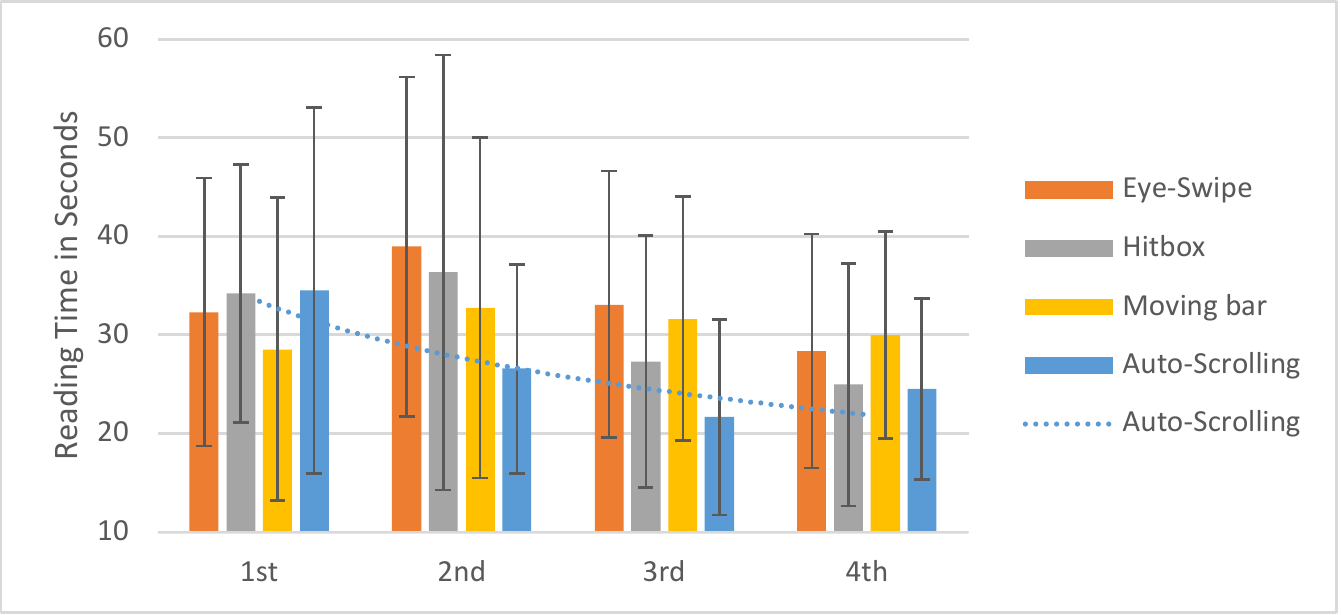}
    \caption{Reading Time Per Page}
    \label{fig:seqpageturn}
    \end{subfigure}
    \caption{Learning effect on gaze scrolling techniques}
\end{figure*}

We present the learning effect in SUS, NASA-TLX, and RTPP in Fig~\ref{fig:seqnasa}, ~\ref{fig:seqsus} and ~\ref{fig:seqpageturn}; that is, the scores in usability and cognitive load, and reading time per page on the order of each gaze interface being introduced. For the learning effect on NASA-TLX and SUS, we run a two-way repeated measures ANOVA with scrolling techniques and experiment order of gaze technique as the two independent variables. The results show that the order has a significant effect on NASA-TLX ($\it{F}(3, 48) = 2.988, \it{p} = .040$), but no significant effect on SUS ($\it{F}(3, 48) = 2.224, \it{p} = .097$). The post hoc pairwise t-test on NASA-TLX suggests that there is significant difference when a gaze interface is introduced first or after learning two gaze interfaces ($\it{t(19)} = 2.60, \it{p} = .017$).

For learning effect on RTPP of mobility condition, we run a three-way repeated measures ANOVA with scrolling techniques, mobility conditions, and  experiment order of gaze technique as the three independent variables. We found the order has a significant effect on RTPP ($\it{F}(3, 49.34) =  5.350, \it{p} = .003$), and there is an interaction effect between order and mobility conditions ($\it{F}(3, 683.85) = 12.844, \it{p} < .001$). 

The post hoc pairwise $\it{t}$-test with Bonferroni correction is conducted on assessing the order effect of each gaze scrolling technique; that is, whether there is a difference when a technique is introduced earlier or later. The results showed significant differences in RTPP between the following order pairs $(1^{st}, 3^{rd})$ ($\it{t(418)} = 3.68, \it{p} = .002$), $(1^{st}, 4^{th})$ ($\it{t(353)} = 4.53, \it{p} < .001$), $(2^{nd}, 3^{rd})$ ($\it{t(362)} = 3.65, \it{p} = .002$),  $(2^{nd}, 4^{th})$ ($\it{t(341)} = 4.40, \it{p} < .001$).

Similarly, the orders of gaze scrolling techniques have a significant impact on the participants' reading time. The participants spend longer on reading when first using a gaze scrolling technique. The possible reason is that the participants are unconsciously distracted by the gaze interface and need more time to learn. The reading time decreases when using the gaze scrolling techniques subsequently.

\section{Discussion}\label{sec:discussion}
Here we reflect on what we have learnt through the user study and identify further research directions. 

\subsection{Real-time Gaze-assisted Scrolling}
Our work presents a collection of real-time gaze-assisted scrolling techniques for a reading application. Our approach is different from the existing gaze-assisted scrolling techniques in Section 2 in terms of gaze estimation techniques, types of gaze interfaces, and the testing platform. More specifically, we take advantage of front camera of a smart phone, and adopt appearance-based gaze estimation to perform real-time gaze estimation. We explore a wider range of explicit gaze interaction methods from dwell-time and pursuit to gaze gesture, and implicit gaze interaction method, the \textit{Auto-Scrolling}. Our reading application is deployed on a phone, which is much smaller than a desktop screen used in the previous studies and thus has higher requirement on the precision of gaze estimation. More importantly, we assess our gaze interfaces under both stationary and mobile conditions, which can shed light on whether gaze interfaces can support real-time interaction for real-world mobile applications, and if so, what type of interaction performs better.

\subsection{Limitation of Our Study} 
The main limitation of our work is that the gaze estimation model is deployed on a server, which can incur extra latency and  raises privacy concerns. To address these issues, our future work will focus on moving the gaze estimation model to the device and performing on-device inference. However, the computation cost can be high so we will look into approaches to reduce computational footprint and optimise battery usage. Also comparisons with other gaze-assisted scrolling techniques will be further explored.

In terms of our user study, our recruited participants had relatively low English proficiency levels (3.2 out of 5), which may have affected their reading speed and patterns. Future studies should include participants who can read in their native languages. In terms of experiment procedure, touching-based scrolling is always the first tested, which may have introduced bias into the study due to counterbalanced issue. In the future, the counterbalanced order of all the scrolling techniques will be applied after a warm up training session. 

\subsection{Impact of Mobility and Environment on Gaze Interaction}
Our results have shown that \textit{Eye-Swipe} and \textit{Hitbox} can perform well under the dynamic condition.
For \textit{Eye-Swipe}, the system only needs to observe a vertical trajectory from the bottom to the top; while for \textit{Hitbox}, a short period of fixation on the bottom area is sufficient. From the semi-structured interview, some participants reflect that \textit{Hitbox}, dwell-time based technique is easier to control while walking. \textit{Auto-Scrolling} is still under-performing due to irregular reading patterns under walking, which is caused by inaccurate gaze estimation due to the changing holding postures and also natural, occasional gaze out of the screen area for checking the walking path~\cite{warlop2020gaze}. 

Mobility makes it challenging to use demanding gaze interfaces like \textit{Moving bar} that requires the execution of complicated gaze actions in an accurate manner. Users can experience too many abortions of the action due to sporadic deviation of gaze. Participants have commented on it as ``tiring'' and ``complicated''.  

Mobility not only changes the holding postures, but also aggravates ``Midas Touch Problem''.  Our user study is conducted in an uncluttered, controlled environment, and each participant is asked to walk on an anticipation-based route. They will only use peripheral vision while walking but occasionally we observe that they will still direct their gaze to checking obstacles and plan paths~\cite{jovancevic2009adaptive, higuchi2013visuomotor, warlop2020gaze, cinelli2009behaviour}. This is a natural behaviour while people walk and is also reflected in the longer reading time while walking compared to sitting, presented in Figure~\ref{fig:pageturnposture}. In a cluttered environment, users have no anticipation-based path and will have more irregular gaze patterns~\cite{hollands2001coordination, warlop2020gaze, higuchi2013visuomotor, jovancevic2009adaptive, cinelli2009behaviour}. During walking, trajectory and dwell can still be captured and have enough tolerance for unstable gaze points; for example,  ballistic trajectories and dwell fixed in one area are more easily distinguished from noisy raw gaze data.

\subsection{Further Improvement on Gaze Estimation}
Appearance-based gaze estimation needs to be more accurate under the mobile condition. First of all, the gaze estimation model is trained on the data collected under the stationary condition, so the model is not tolerant to various distances and tilting angles between the screen and face. Collecting more data under a diversity of dynamic conditions has already been recognised as an important direction. 

Calibration works even for a short 5-second calibration using a lightweight SVR. However, when participants change holding postures, the model becomes less accurate. It is desirable to perform continuous calibration. The key will be on when and how to perform implicit calibration; that is, without interrupting the current task, could we find a surrogate or proxy to get the true points?

\subsection{More Thoughts on Implicit Gaze}
\textit{Auto-Scrolling}, implicit gaze interface, is not well perceived in our experiment. There are various reasons behind that; for example, our reading speed prediction is a simple regression algorithm and only takes a few nonconsecutive gaze points at the top and middle regions of the screen.

\textit{Auto-Scrolling} can work well on some participants who pay high attention on the reading material and reads in a continuous, regular pace. They comment it as ``good for reading novels'', ``very natural to turn the pages as I read for fun'', and ``fit my reading habits very well''. By examining this participant's data, we can see that his eye movements cover the whole page and the transition of gaze from the top to the bottom is clear. However, it does not work for most participants when they do not have linear reading patterns. For example, one participant comments ``while reading the poem, the technique prevented me from thinking.'' Readers may divert their gaze out of the screen or fixate at some particular words for thinking or mind-wandering. This adds extra complication and ways to improve better prediction include using more gaze points, using extra information about the reading material such as textual features to infer users' preference, and adopting a more sophisticated prediction algorithm.

\section{Conclusion and Future Work}\label{conclusion}
In this study, we evaluated four gaze interaction methods for scrolling pages in a mobile reading application through a user study and a detailed empirical analysis. Overall, explicit gaze interaction methods perform well under the mobile condition, and are consistently ranked higher as user preferred techniques than touch-based scrolling. Participants found explicit gaze interaction methods enjoyable and easy to use and and felt they maintained control over their reading pace. Gesture and Dwell-time based gaze interaction methods proved more robust than Pursuit during walking. 
This is a promising result, confirming the usability of gaze in mobile applications and encouraging further innovative designs on handheld mobile devices.

\begin{acks}
We thank all of our participants for their time and contribution to the project. We thank Shijing He for his valuable support on paper figure design and Jason Jacques, Kenneth Boyd and Angela Miguel for their constructive feedback and comments. And Lei, Y. and Wang, Y. acknowledge the financial support by the University of St Andrews and China Scholarship Council Joint Scholarship.
\end{acks}

\bibliographystyle{ACM-Reference-Format}
\bibliography{main}

\received{November 2022}
\received[revised]{February 2023}
\received[accepted]{March 2023}

\end{document}